\documentstyle[psfig]{mn}

\newif\ifAMStwofonts

\ifoldfss
  \ifCUPmtlplainloaded \else
    \NewTextAlphabet{textbfit} {cmbxti10} {}
    \NewTextAlphabet{textbfss} {cmssbx10} {}
    \NewMathAlphabet{mathbfit} {cmbxti10} {} 
    \NewMathAlphabet{mathbfss} {cmssbx10} {} 
  \fi
  \ifAMStwofonts
    \ifCUPmtlplainloaded \else
      \NewSymbolFont{upmath} {eurm10}
      \NewSymbolFont{AMSa} {msam10}
      \NewMathSymbol{\upi}     {0}{upmath}{19}
      \NewMathSymbol{\umu}     {0}{upmath}{16}
      \NewMathSymbol{\upartial}{0}{upmath}{40}
      \NewMathSymbol{\leqslant}{3}{AMSa}{36}
      \NewMathSymbol{\geqslant}{3}{AMSa}{3E}

       \let\le=\leqslant
       
    \fi
  \fi
\fi 

\ifnfssone
  \newmathalphabet{\mathit}
  \addtoversion{normal}{\mathit}{cmr}{m}{it}
  \addtoversion{bold}{\mathit}{cmr}{bx}{it}
  \newmathalphabet{\mathbfit} 
  \addtoversion{normal}{\mathbfit}{cmr}{bx}{it}
  \addtoversion{bold}{\mathbfit}{cmr}{bx}{it}
  \newmathalphabet{\mathbfss} 
  \addtoversion{normal}{\mathbfss}{cmss}{bx}{n}
  \addtoversion{bold}{\mathbfss}{cmss}{bx}{n}
  \ifAMStwofonts
    \ifCUPmtlplainloaded \else
      \UseAMStwoboldmath
      \makeatletter
      \new@mathgroup\upmath@group
      \define@mathgroup\mv@normal\upmath@group{eur}{m}{n}
      \define@mathgroup\mv@bold\upmath@group{eur}{b}{n}
      \edef\UPM{\hexnumber\upmath@group}
      \new@mathgroup\amsa@group
      \define@mathgroup\mv@normal\amsa@group{msa}{m}{n}
      \define@mathgroup\mv@bold\amsa@group{msa}{m}{n}
      \edef\AMSa{\hexnumber\amsa@group}
      \makeatother
      \mathchardef\upi="0\UPM19
      \mathchardef\umu="0\UPM16
      \mathchardef\upartial="0\UPM40
      \mathchardef\leqslant="3\AMSa36
      \mathchardef\geqslant="3\AMSa3E

       \let\le=\leqslant

    \fi
  \fi
\fi 

\ifnfsstwo
  \DeclareMathAlphabet{\mathbfit}{OT1}{cmr}{bx}{it}
  \SetMathAlphabet\mathbfit{bold}{OT1}{cmr}{bx}{it}
  \DeclareMathAlphabet{\mathbfss}{OT1}{cmss}{bx}{n}
  \SetMathAlphabet\mathbfss{bold}{OT1}{cmss}{bx}{n}
  \ifAMStwofonts
    \ifCUPmtlplainloaded \else
      \DeclareSymbolFont{UPM}{U}{eur}{m}{n}
      \SetSymbolFont{UPM}{bold}{U}{eur}{b}{n}
      \DeclareSymbolFont{AMSa}{U}{msa}{m}{n}
      \DeclareMathSymbol{\upi}{0}{UPM}{"19}
      \DeclareMathSymbol{\umu}{0}{UPM}{"16}
      \DeclareMathSymbol{\upartial}{0}{UPM}{"40}
      \DeclareMathSymbol{\leqslant}{3}{AMSa}{"36}
      \DeclareMathSymbol{\geqslant}{3}{AMSa}{"3E}

       \let\le=\leqslant

    \fi
  \fi
\fi 

\ifCUPmtlplainloaded \else
  \ifAMStwofonts \else 
    \def\upi{\pi}
    \def\umu{\mu}
    \def\upartial{\partial}
  \fi
\fi

\begin{document}

\title[
Constraints on Merger Models from Globular Cluster Systems
]
{
Constraints on the Merger Models of Elliptical Galaxies from 
their Globular Cluster Systems
}

\author[
Markus Kissler-Patig, Duncan A. Forbes, Dante Minniti
]
{
Markus Kissler-Patig$^{1,2}$, Duncan A. Forbes$^3$, Dante Minniti$^{4,5}$\\
$^1$Lick Observatory, University of California, Santa Cruz, CA 95064, USA\\
(E-mail: mkissler@ucolick.org)\\
$^2$Feodor Lynen Fellow of the Alexander von Humboldt Foundation\\
$^3$School of Physics and Astronomy, 
University of Birmingham, Edgbaston, Birmingham B15 2TT \\
(E-mail: forbes@star.sr.bham.ac.uk)\\
$^4$ Lawrence Livermore National Laboratory, Livermore, CA 94550, USA\\
(E-mail: dminniti@igpp.llnl.gov)\\
$^5$ Departamento de Astronom\'\i a y Astrof\'\i sica, P.~Universidad
Cat\'olica, Casilla 104, Santiago 22, Chile
}

\pagerange{\pageref{firstpage}--\pageref{lastpage}}
\def\LaTeX{L\kern-.36em\raise.3ex\hbox{a}\kern-.15em
    T\kern-.1667em\lower.7ex\hbox{E}\kern-.125emX}

\newtheorem{theorem}{Theorem}[section]

\label{firstpage}

\maketitle

\begin{abstract}
The discovery of proto--globular cluster candidates in many current--day mergers allows 
us to better understand the possible effects of a merger event on the globular cluster 
system of a galaxy, and to foresee the properties of the end--product. By 
comparing these expectations to the properties of globular cluster systems 
of today's elliptical galaxies we can constrain merger models. The observational
data indicate that {\it i}) every gaseous merger induces the formation of new 
star clusters, {\it ii}) the number of new clusters formed in 
such a merger increases with the gas content of the progenitor galaxies.
Low--luminosity (about $M_V>-21$), disky ellipticals are generally thought to 
be the result of a gaseous merger. As such, new globular clusters are expected 
to form but have 
not been detected to date. We investigate various reasons for the 
non--detection of sub--populations in low--luminosity ellipticals, i.e.~absence of 
an old population, absence of a new population, destruction of one of the 
populations, and finally, an age--metallicity conspiracy that allows old and new 
globular clusters to appear indistinguishable at the present epoch. All of these 
possibilities lead us to a similar conclusion, namely that low--luminosity ellipticals did 
not form recently ($z<1$) in a gas--rich merger, and might not have formed in a
major merger of stellar systems at all. 
High--luminosity ellipticals do reveal globular cluster sub--populations.
However, it is difficult to account for the two populations in terms of
mergers alone, and in particular, we can rule out scenarios in which the
second sub--population is the product of a recent, gas--poor merger.

\end{abstract}



\section{Introduction}

Globular clusters are thought to be good tracers of the past evolution of 
their parent galaxy. Recreating the history of a galaxy from the current 
properties of its globular cluster system is complex and challenging. However,
if a specific event is expected in the history of a galaxy, such as a major
merger, and this event is supposed to leave a clear signature in the
globular cluster system (e.g.~the formation of a new population
of globular clusters), then studies of globular cluster systems can be used to
confirm or rule out the occurrence of such events.

In this paper, we examine the constraints that can be set from the
current observations of globular cluster systems on the merger model for
elliptical galaxies. The main challenge is to understand the apparent
presence of {\it only one} population of globular clusters in the 
luminosity and colour distributions of small ellipticals.
Ashman \& Zepf (1992) made a variety of predictions
for the properties of globular cluster systems after a merger of two
gas--rich galaxies. One of their main predictions is that mergers
will leave a clear signature in the form of a {\it second} population of 
globular clusters, formed during the merger,  which is added to 
the old populations of globular clusters brought in by the progenitors.
A notable success of their model is the presence
of young proto--globular cluster candidates in currently merging
systems, confirming the basic idea that such events can modify the
existing globular cluster systems. Furthermore some of their 
predictions seem to be verified for the globular cluster systems of giant
elliptical galaxies (Zepf \& Ashman 1993): two or more globular cluster
populations are detected in colour distributions, the red (presumably
new) globular clusters are more concentrated toward the center of the
galaxy as expected if these formed from the in--falling gas. 

However, whether the globular cluster systems of all ellipticals verify 
these predictions remains an open question. The
properties of globular clusters in small and large ellipticals seem to differ.
Studies of systems in small ellipticals failed to detect more than one
population of globular clusters (Kissler-Patig 1997a). Moreover, Forbes, Brodie \& Grillmair (1997) 
have recently argued that the observational data on globular clusters in 
high--luminosity ellipticals (i.e.~gE and cD galaxies) cannot be explained
in detail by an Ashman \& Zepf type merger, and could be explained in
other formation scenarios. 

We adopt a working definition of --18 $>$ M$_V$ $>$ --21 for low--luminosity 
and --21 $>$ M$_V$ $>$ --23.5 for high--luminosity ellipticals (for H$_0$ =
75 km s$^{-1}$ Mpc$^{-1}$) throughout this paper. Further, we will refer
to the {\it new} population of globular clusters as the population formed
in the merger event, and to the {\it old} population of globular clusters 
as the ``original'' population, present before any merger event. 

We examine the numerous observations of proto--globular
clusters in merging galaxies to extract the signature that these events
will leave on the globular cluster systems (Sect.~2). In particular, we
examine if, and how many, globular clusters are expected to form in the
different (gas--rich/gas--poor) mergers.
In Section 3, we consider low--luminosity and high--luminosity galaxies
in turn. We investigate the various alternatives that could explain the
presence of only one population of globular clusters in low--luminosity
ellipticals, despite the fact that they are generally thought to have formed 
several Gyr after a gaseous merger event and thus should show a second
population of globular clusters. 
The most interesting alternative (Sect.~3.2) is the presence of two
populations whose mean ages and metallicities conspire to let them appear
the same in colour and magnitude distributions.
The various alternatives allow us to put constraints on the occurrence of a 
merger event in the recent history of the parent galaxy. We then examine
in Sect.~3.4 if similar arguments can also constrain the history of
high--luminosity ellipticals, and in Sect.~3.5 if additional constraints
can be obtained from the properties of ongoing mergers. We summarize our 
findings and conclusions in Sect.~4.

 
\section{Our present knowledge}

In the following section, we use the recent studies of globular clusters in
mergers to derive two important facts: new globular clusters will form
in a gas--rich merger event, and their number will roughly scale with
the amount of gas involved. Further, we summarize briefly the current   
scenarios of early--type galaxy formation by mergers. 

\subsection{The formation of globular clusters in gaseous mergers}

Proto--globular clusters have been observed in all recent merger
systems studied to date: e.g.~NGC 3597 (Lutz 1991, Holtzman et al.~1996), 
NGC 1275 (Holtzman et 
al.~1992), NGC 7252 (Whitmore et al.~1993, Schweizer \& Seitzer 1993), 
NGC 4038/4039 (Whitmore \& Schweizer 1995), NGC 5128
(Minniti et al.~1996), NGC 5018 (Hilker \& Kissler-Patig 1997), NGC 3921 
(Schweizer et al.~1996), NGC 3256 (Zepf et al.~1997), NGC 6052
(Holtzman et al.~1996), NGC 3610  (Whitmore et al.~1997).

Further, the number of newly formed globular clusters seems to vary with the
type of the progenitor galaxies. During the collision of two gas--rich galaxies 
such as NGC 4038/4039 (two Sb/Sc spirals), a very large number of  
globular clusters have formed. Whitmore \& Schweizer (1995) report 700 potential
new globular clusters, while the spiral progenitors might have had a few 
hundred each, i.e.~around 100\% new globular clusters (by which we mean
as many as old ones) might have formed. 
Further examples are NGC 7252 (the merger of two massive Sc galaxies,
Fritze-v.Alvensleben \& Gerhard 1994) and NGC 3610 (also a good
candidate for a disk/disk merger, Schweizer \& Seitzer 1992)
formed 40\% and 70\% new globular clusters compared to the existing ones
(Whitmore et al.~1997). A large
number of new globular clusters is also seen in the case of NGC 3256 (Zepf
et al.~1997), which is rich in molecular gas.  

An example of a merger between a gas--rich (Sc)
and an S0/Sa galaxy is thought to be 
NGC 3921 (Schweizer et al.~1996). In this case the number of new
globular clusters lies around 40\% of the old globular cluster population. Going to even earlier
types as for NGC 5128 or NGC 5018 (both of which may be the result of a disk system
falling onto a gas--poor early--type galaxy), the number of new globular clusters
is even smaller. In NGC 5128 the ``intermediate'' age globular clusters in the inner
region represent about 20\% of the old population (Minniti et al.~1996),
however, since young clusters form preferentially in the center this
might be an upper limit. In NGC 5018, Hilker \& Kissler-Patig (1996)
estimate that the new globular clusters represent at most 10\% of the old
population. NGC 1275 might be a peculiar case, but here also a disk
system is falling onto a early--type galaxy, and while the estimated old
population is of several thousands globular clusters (N{\o}gaard--Nielsen et
al.~1994), the number of new clusters appears to be small (Holtzman et
al.~1992).

In summary, the observations indicate that: {\it i)} 
in a dissipational merger (i.e.~one involving gas) new globular clusters will
always form. Thus galaxies that underwent a gaseous merger event
at any epoch {\it must} have formed a population of globular clusters 
associated with that merger event. {\it ii)} The number of new globular 
clusters produced will be
proportional to the available gas content. Although there will be
other factors involved (e.g. gas density, collision velocity) we
expect mergers that involve a large amount of gas to create a large 
number of new globular clusters. A general assumption is that
the proto--globular clusters seen today will indeed evolve into globular
cluster like objects after several Gyr (see Sect.~3.2 and 3.5 for a more
detailed discussion of this point).

\subsection{The formation of early--type galaxies in mergers}

Faber et al.~(1997, see also e.g.~Bender 1997) have recently summarized the 
properties of hot galaxies 
within the framework of hierarchical clustering and merging. They
concluded that low--luminosity, disky 
ellipticals seem generally compatible with formation in 
dissipative, gas--rich mergers. Kauffmann (1996) further proposed a dependence 
on environment, in the sense that intermediate--luminosity galaxies in clusters 
are old, while those in the field were mostly formed by more recent mergers of 
spirals. Recently, De Jong \& Davies (1997) found a correlation between
the H$\beta$ absorption index, indicative of age, and the isophotal
shape of elliptical galaxies, in the sense that disky ellipticals are
younger than boxy ones or had a small amount of recent star formation.  

Boxy, high--luminosity ellipticals on the other hand are thought to have formed 
in gas--poor mergers. Their history can be divided in two parts: an
early--phase of merging of the largely unknown progenitors,
and a late phase of accretion involving small companions. Further, there is
increasing evidence that these galaxies have formed the bulk of their stars at
high redshifts, i.e.~$z\sim 3$ (e.g.~discussion in Renzini 1997).

Although there is not a well--defined separation between disky and boxy
galaxies, the transition occurs around M$_V = -21$ (e.g.~Bender et
al.~1989). 


\section{Are the properties of globular cluster systems compatible with merger 
scenarios?}

Combining the results from the previous section, we expect small
ellipticals, if they formed relatively recently in
dissipational mergers, to have formed new globular clusters in addition
to the old ones from the progenitors.
We focus on these galaxies in Sect.~3.2 and discuss the different 
possibilities that would explain why only one population is seen today.
We then examine what constraints this puts on the merger history. We will 
discuss the large ellipticals in Sect.~3.4. In particular, to see if 
such constraints are also valid in their case. We 
start by revisiting some arguments that have been put forward in favour
or against mergers of spirals creating ellipticals, using the properties 
of the globular cluster systems.

\subsection{Old problems revisited}

The properties of globular cluster systems in early--type galaxies have been 
described in 
various reviews (e.g.~Harris 1991, Richtler 1995, Ashman \& Zepf 1997). 
Some of these properties have been used by
van den Bergh (1990) to argue against mergers. In particular,
ellipticals have more globular clusters per unit starlight
(i.e.~a higher specific frequency) than spirals. It is unclear
whether this could be overcome by the creation of new globular clusters 
in the merger (e.g.~Ashman \& Zepf 1992) or not (e.g.~Harris 1994, Forbes,
Brodie \& Grillmair 1997). This depends on the unknown ratio of
globular clusters versus star formation efficiency in mergers.
Several new compilations (Harris 1996, Kissler-Patig 1997a, Ashman \& Zepf 1997)
show that the specific frequency discrepancy between spirals and
low--luminosity ellipticals is actually small, if existent at all, 
when the different 
mass--to--light ratios of spirals and ellipticals are taken into account 
(low--luminosity ellipticals have typically 200--500 globular clusters,
similar to spirals of comparable mass). However, large differences remain
when comparing spirals with high--luminosity ellipticals.
So the problem 
of too many globular clusters present in bright ellipticals may still exist, 
but we stress that alternative explanations to the merger picture could 
overcome this problem (e.g.~the summary in Forbes, Brodie \& Grillmair 1997).

Another controversial point is the following. Recently
it has become clear that several high--luminosity ellipticals
have broad, multi--modal globular cluster colour distributions (e.g.~Lee \& 
Geisler 1993; Geisler, Lee \& Kim 1996; Forbes, Brodie \& Huchra 1997). 
This multi--modality indicates the presence of different globular cluster
sub--populations, i.e.~several epochs or mechanisms of formation. 
In the literature summary included in Kissler-Patig (1997a) and 
Forbes, Brodie \& Grillmair (1997), only {\it high--luminosity} (M$_V<$--21) 
ellipticals revealed obvious bimodal globular cluster colour distributions
(we will come back to this point in the next section). Moreover,
typically the two peaks
contain equal numbers of red and blue globular clusters within a factor two, 
which is contrary to the expectations from a gas--poor merger event.
From observations of ongoing mergers (Sect.~2.1) we would expect the number 
of red clusters to be of the
order of 10\% or less, i.e.~the ratio of red to blue clusters to be
$<0.1$, if the gas--poor merger was the only process responsible for the 
formation of the red globular clusters.
Forbes, Brodie \& Grillmair (1997) have concluded that the
bimodality of the globular cluster colour distributions in the high--luminosity 
ellipticals is better explained by a multi--phase collapse than by a merger 
event. Recently, Kissler-Patig et al.~(1998) obtained spectra for 
globular clusters in NGC 1399 (a galaxy with a multi--modal globular
cluster distribution) and concluded from the absorption line
indices, that all blue and most red clusters were old, but {\it a very small 
fraction}
of extremely red clusters were much more metal rich and could be younger,
i.e.~they may
have formed in a later merger. 
Cohen, Blakeslee \& Rhyzov (1998) obtained spectra for globular clusters in
M87 and came to the similar conclusion that most globular clusters around this
galaxy are old.
Therefore, care should be taken when reaching
conclusions from multi--modal colour distributions. Their absence certainly
hints at the absence of sub--populations (see below), but the
presence of sub--populations does not automatically imply a recent (or even
past) merger, nor does it exclude it (see Sect.~3.4).

Another interesting point is that globular cluster system properties
seem to differ between low--luminosity ellipticals, and high--luminosity
ellipticals, and that two classes may be preferred to a
continuous relation (Kissler-Patig 1997a). It seems therefore misleading
to discuss the properties of globular clusters in ellipticals without
differentiating between low-- and high--luminosity galaxies.
In particular, the globular cluster systems in low--luminosity
ellipticals do not obviously show the properties first predicted by Ashman
\& Zepf (1992) for systems resulting from a recent merger of two spirals.
Since these two classes of ellipticals are thought to have different
formation histories from various independent lines of evidence
(see Sect.~2.2), it is worth examining them in turn.

\subsection{Low--luminosity ellipticals}

We focus on the low--luminosity ellipticals which, as mentioned above,
are generally thought to be the result of a gaseous merger and so should 
have formed many new globular clusters. One also expects such
ellipticals to show 
the old globular clusters associated with the progenitor galaxies.
These two globular cluster populations (one new and one old) should show up 
in the globular cluster luminosity and colour distributions of the final 
system. Small ellipticals are therefore ideal
candidates to search for the signature of a merger event in their
globular cluster system.

Interestingly, the properties of the globular cluster systems in 
small ellipticals (see Kissler-Patig 1997a) do not clearly imply the presence 
of sub--populations.
There is no known case with a clear bimodal colour distribution or a colour
gradient in the
globular cluster system of a low--luminosity elliptical. Further, the globular 
clusters are not more extended, but follow the stellar 
light distributions (in radial density, ellipticity, position angle),
despite the prediction that after a merger, the old population should be
more diffuse than the newly formed stars and globular clusters.
From the properties of their globular cluster systems, 
low--luminosity ellipticals do not, a priori, require a recent merger event.
The predictions of the simple Ashman \&
Zepf (1992) merger scenario (implying the formation of a new globular
cluster population) have yet to be seen in the globular cluster systems of any
low--luminosity ellipticals.

One disadvantage of studying low--luminosity ellipticals is
that they tend to have fewer globular clusters in total, making any colour 
bimodality more difficult to observe. But, if hidden by small number 
statistics, we might expect the colour distributions to be on average as broad as
those for large ellipticals. However, this does {\it not} appear to be the
case (Kissler-Patig et al.~1997a) with small ellipticals having
distributions that are 3 to 4 times narrower than the ones in large
ellipticals.

In summary, gas--rich mergers would imply the formation of a large number of
new globular clusters, in addition to the existing old 
globular clusters associated with the progenitor galaxies. One expects to see
these two populations of globular clusters in low--luminosity ellipticals, 
and yet they are not seen. 
There are several possibilities to explain this apparent absence of
sub--populations:

\begin{itemize}
\item{only old globular clusters are present, as the new globular clusters 
were rapidly destroyed,}
\item{or only new globular clusters are present because the progenitor 
galaxies had very few old globular clusters,}
\item{or only new globular clusters are present because old globular clusters 
have been preferentially destroyed or removed from the galaxy,}
\item{or only old globular clusters are present as very few were created 
in the merger,}
\item{or new and old globular clusters are present but are essentially 
indistinguishable in magnitude and colour.}
\end{itemize}

We discuss each point in turn.

$\bullet$ The rapid destruction of all newly formed globular clusters  
seems unlikely given the observations of similar masses for the
new and old globular clusters (see Sect.~3.2.1). Elmegreen \& Efremov
(1997) summarized recent observations and argued that newly formed
globular clusters in mergers resemble more closely open clusters in
their formation process {\it but} are much more massive and more compact
than the latter, and have an initial mass distribution similar to that of 
the old globular clusters, i.e.~new and old clusters will be indistinguishable 
after several Gyr. Destruction processes would be expected to affect both
populations. The new globular clusters are expected to be more
spatially concentrated and therefore suffer preferentially from efficient
destruction processes such as bulge shocking (Gnedin \& Ostriker 1997).
But such processes need of the order of a Hubble time to destroy a
significant fraction of the globular clusters. Destruction processes 
would have had a longer time to act on the old globular 
clusters. Further, in the Elmegreen \& Efremov picture destruction would be 
less efficient for the more compact young objects, eliminating if anything, 
more old globular clusters. Finally, the fact that new globular 
clusters survive at least a
few Gyr is directly demonstrated by observations of ``young'' globular 
clusters in Gyr old mergers (e.g.~NGC 3610, NGC 5018, NGC 5128, NGC 7252). 
It appears therefore unlikely that a large population of newly formed 
globular clusters would be rapidly and efficiently destroyed without the old 
population 
being much affected, unless the new clusters turned out to be very different
from old globular clusters (see Sect.~3.5).

$\bullet$ Another possibility for the absence of sub--populations in 
low--luminosity ellipticals might be that today's population of globular 
clusters  formed {\it entirely} in
the merger event, i.e.~little or no old globular clusters have been brought
in by the progenitors. 
However, {\it all} current galaxies (late-- or early--type) larger than M$_V 
\simeq -15$ are known to host a population of globular clusters. Even Sd
galaxies, with presumably the lowest globular cluster to mass ratio 
(e.g.~Harris 1991), would, when scaled up to masses able to produce a
low--luminosity elliptical, bring a few hundred clusters with them, and
produce at most around a thousand new ones, if NGC 4038/4039 (a late--type
galaxy merger, see Sect.~2.1) is taken as reference. The progenitor
galaxies must have been unevolved gas disks if they did not
bring with them globular clusters into the final system.
Further, since the globular clusters
in low--luminosity ellipticals are thought to be old from their colours and
magnitudes, any merger that formed them must have happened at an early epoch.
Low--luminosity ellipticals would have formed before or only shortly after
the main epoch of globular cluster 
and star formation, from mainly gaseous progenitors. The
globular clusters seen today would be the ``new'' ones in the sense that
they were formed in the merger, but they would be about 15 Gyr old.

$\bullet$ Another explanation for the presence of only new globular clusters
could be the preferential destruction or removal of the old globular
clusters. The preferential destruction of old globular clusters is unlikely
(see the first point). However, a preferential removal of the old globular
clusters, which are spatially more extended around the galaxy, is not excluded.
Muzzio (1987) summarized the
effects of stripping and harassment in clusters of galaxies on globular
cluster systems. Although the exact effects depend on the
characteristics of environment, galaxy, and globular cluster system, tidal
stripping is not negligible and will affect primarily the outer
globular clusters. This could explain the absence of a large blue
population in some ellipticals, but this needs a more detailed analysis to
estimate its importance. The consequence would be similar to the previous
point, i.e.~we would see only the ``new'' globular clusters, but they 
formed long ago.

$\bullet$ The absence of new globular clusters may
simply mean that they never formed or that they are very few in number. 
Either way, given the observational situation described in Sect.~2.1, this
would rule out that low--luminosity elliptical formed by
major gaseous merger events.

$\bullet$ Yet another option could be that new globular clusters are missed in 
these galaxies, because they neither differ in luminosity nor colour from 
the old globular clusters.
This would have to be the result of a conspiracy between mass, age, and
metallicity of the various sub--populations. In the following, we investigate 
in more detail the allowed range for these parameters, that will allow  new 
globular clusters to appear similar in luminosity and colour to old, 
metal--poor globular clusters. 

\subsubsection{Hiding a second population: Constraints from the similar luminosity 
functions}

We first examine the conditions necessary to obtain similar 
luminosity functions for the new and old globular clusters.
We then derive the constraints on age and metallicity to reproduce the
current observations.

The luminosity function of old globular clusters in all well--studied systems
are observed to be single peaked, with 
a ``universal'' absolute turn--over magnitude $M_V^{TO}\simeq -7.5$. This 
turn--over magnitude depends only weakly on galaxy luminosity and type
(Harris 1996). 
The difference in the peak of the luminosity function between different 
globular cluster populations must be smaller than the detection limit, that we estimate to be 
$\le$ 0.4 mag in the current data sets (e.g.~Forbes et al.~1996, 
Kissler-Patig et al.~1997b, Forbes, Brodie \& Huchra 1996). 
Can new and old globular clusters have similar luminosities? 
In general, the luminosities of globular clusters can vary 
{\it i)} with the mass of the objects if a constant mass--to--light
ratio (M/$L$) is assumed, {\it ii)} with M/$L$ if a constant globular cluster 
mass distribution is assumed or in other words with the stellar initial mass 
function (IMF) within the clusters, {\it iii)} with the cluster ages,
and {\it iv)} with the cluster metallicities.  

{\it i) Similar mass distributions:}

Will new and old globular clusters have comparable mass distributions
to allow similar luminosity functions?
Meurer (1995) first argued that newly formed globular clusters 
in NGC 4038/4039 have, at the observed bright end, a magnitude distribution 
that is compatible with that observed for old globular clusters in galaxies,
i.e.~the underlying globular cluster mass distributions must be similar. 
Schweizer et al.~(1996) confirmed in NGC 3921 that the luminosity function of
the young objects followed a power--law, but also pointed out that it does
so down to lower masses than expected.
They pointed out that this luminosity function will only evolve into what
is currently seen for old globular clusters, if globular clusters are
preferentially destroyed by processes acting more efficiently at the
low--mass end. Recently, Elmegreen \& Efremov (1997)
argued that the star cluster formation mechanism was universal, and 
explicitly that old globular clusters and younger ones (formed in mergers)
have identical initial mass distributions. As time evolves,
destruction processes will make the luminosity function look like the one
observed today in nearby galaxies. This was recently  supported
observationally by Harris, Harris \& McLaughlin (1998) who found no significant 
difference in the luminosity functions of the two old populations in M87
and concluded that the destruction processes at the low--mass end do not
affect the bright end and turn--over of the luminosity function.
We will therefore assume that two populations will be
indistinguishable by their mass distribution, down to and slightly
beyond the turn--over of the luminosity function, after several Gyr.

{\it ii) Similar mass--to--light ratios:}

The mass--to--light ratio of a globular cluster changes with age but is
predictable by population synthesis models (see below) if the IMF of the 
globular cluster is known. Are large variations
of the IMF expected between the new and the old globular clusters?
Dubath \& Grillmair (1997) showed that globular clusters in the Milky Way, M31, Fornax, 
the Magellanic Clouds, and Cen A have extremely similar $M/L_V$ ratios,
after correcting for the different ages. The result of 
Dubath \& Grillmair indicate that only small variations at the lower end
of the IMF are expected and the mass function of the low--mass stars in 
clusters will be constant. We note that Brodie et al.~(1998) recently found 
evidence from spectroscopy of a slightly flatter than Salpeter IMF for the 
newly formed star clusters in NGC 1275. However, the difference is
mostly noticeable at the high--mass end and while these clusters are young. Once 
these clusters will have aged by a few Gyr, the different $M/L$ between
these clusters and the ones with a Salpeter IMF will be only marginal.

{\it iii) and iv) An age--metallicity conspiracy}

From the two points above, we conclude that no significant variations of the 
mass function of globular clusters and/or the mass--to--light ratio are
expected between young and old globular clusters. This implies further that
neither the mass distribution nor $M/L$ can be significantly 
fine--tuned to compensate for any age or metallicity differences. 
To explain the similar luminosity
functions of the new and old clusters, despite different mean ages and mean
metallicities, {\it we must invoke a age--metallicity conspiracy}.
Such a conspiracy might partly be expected since the luminosity of a globular 
cluster decreases with age and metallicity but new (young) globular
clusters are
expected to be more metal--rich than old ones, since they probably
formed from more enriched material. We quantify this conspiracy in the
next section together with the constraints set by the narrow colour
distributions.

\subsubsection{Hiding a second population:
Constraints from the narrow colour distributions}

If the mean age and metallicity of new and old globular clusters conspire,
we can use the narrow colour distribution observed in low--luminosity
ellipticals to give us another constraint. 
The colour of a globular cluster gets redder with increasing age and increasing
metallicity.
The narrow colour distributions observed in small ellipticals
(e.g.~Kissler-Patig et al.~1997a, Neilsen, Tsvetanov \& Ford 1997),
do not allow all age--metallicity combinations. If we
assume that the four low--luminosity ellipticals in Fornax 
are representative, then the typical
dispersion in a V--I colour distribution (after correcting for 
errors in the photometry) is about $\sigma < 0.1$ mag (Kissler-Patig et 
al.~1997a). 

We have combined the constraints from luminosity and colour
in Fig.~1. The mean V--I colour of a globular cluster population is plotted 
versus its metallicity as a function of age. We start with the observed
single population, then add a second population and vary its mean age
and metallicity until the two populations match both in colour and magnitude 
within the detection limits ($\Delta(V-I)=0.1$ and $\Delta
V=0.4$). We will use the models of 
Worthey (1994) in the following, but obtained very similar results when
using the models of Bruzual \& Charlot (1993, 1997) and of
Fritze--v.Alvensleben \& Burkert (1995). The latter predict even
tighter constraints than derived below.

We start in the upper left panel with the colour distribution typically 
observed in low--luminosity ellipticals, i.e.~with a mean colour around
V--I\ $\simeq$1.0 and a dispersion of about 0.1 mag. We have assumed the
colour distribution of NGC 1427 to be representative (mean 
V--I\ $=1.05\pm0.05$, see also Sect.~3.3).
We associate this mean colour with the old population, and assume an age
of 17 Gyr (the oldest available age in
the Worthey models). The corresponding mean metallicity
of $-1.5<$[Fe/H]$<-0.75$ dex, is in good agreement with spectroscopic
determinations (Kissler-Patig et al.~1998). The dotted lines show the
colour range ($\pm 0.1$ mag) within which a second population would be
indistinguishable in colour from the old population.
In the upper right panel, we now introduce three new populations. One has the
same age as the old population (17 Gyr), the second is half as old (8
Gyr), the third has an age of 5 Gyr . 
Their mean colours vary along the solid lines as a function of 
metallicity (isochrones from Worthey 1994).
The range of colours spanned by the initial old globular cluster population 
is now shown as a hatched 
horizontal band. All new populations match the old one in colour, if
their mean metallicities are approximately [Fe/H]$<-0.3$, [Fe/H]$<-0.1$
and [Fe/H]$<0.0$ dex for 17 Gyr, 8 Gyr and 5 Gyr respectively. 

However, as mentioned above, colour is not
the only constraint and the mean magnitude of these two populations will
change with age and metallicity. This is illustrated in the lower left
panel. As the new 17 Gyr population gets more metal rich, the
globular clusters become fainter. Once [Fe/H]$>-0.25$, their mean
magnitude would differ by more then 0.4 mag in V from the faintest
possible old population (17 Gyr, [Fe/H]$=-0.75$). Both populations could
be disentangled in the globular cluster luminosity function. Similarly,
the 8 Gyr population at metallicities of [Fe/H]$<-0.5$ becomes 
0.4 mag brighter in V than the brightest possible old population (17
Gyr, [Fe/H]$=-1.5$). Such a population could then be identified in the 
globular cluster
luminosity function. These metallicities ([Fe/H]$>-0.25$ for the new 17
Gyr population, [Fe/H]$<-0.5$ for the new 8 Gyr population) are
therefore excluded if the new and old populations are to be
indistinguishable in colour {\it and} luminosity. The 5 Gyr
population shows an extreme case: it can match the colour of the old 
population for metallicities below solar, but would then always be much 
brighter than an old population. For this young age, no combination
of age and metallicity allows the new globular clusters to be hidden
both in colour and luminosity.
The forbidden metallicity ranges due to the constraint from
the luminosity function are marked as dashed lines.

Finally, we show in the lower right panel a larger range of ages and
combine the colour and luminosity constraints to show the narrow 
range of age and metallicity combinations (shaded area) that any new 
population of globular clusters must occupy, in order to be
indistinguishable in current observations from the old one.
The mean age of the new globular cluster population must lie between 8 and 
17 Gyr, and its mean metallicity between $-2.0<$[Fe/H]$<-0.4$ dex.
Furthermore, age and metallicity must tightly conspire (e.g.~8 Gyr old 
globular clusters would have to be have a mean metallicity of 
$-0.5<$[Fe/H]$<-0.1$). 

\begin{figure*}
\psfig{figure=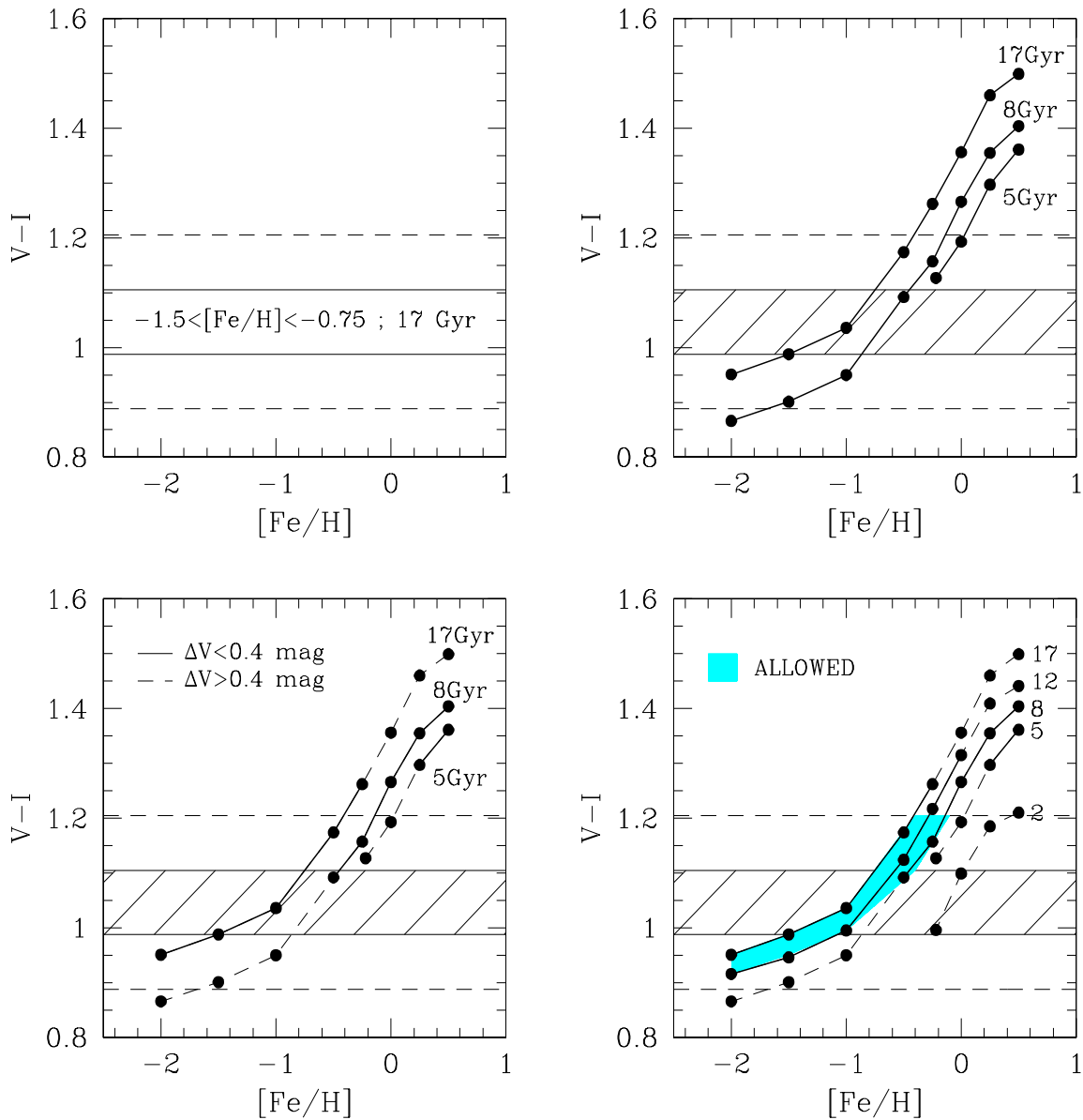,height=16cm,width=16cm
,bbllx=8mm,bblly=57mm,bburx=205mm,bbury=245mm}
\caption{Hiding new globular clusters both in colour and luminosity. The
upper left panel shows the colour distribution of an old globular cluster
population, the dashed lines mark the range within which a new
population would be indistinguishable from the old one in colour. The upper 
right panel shows how colour varies with metallicity for 3 different
new populations (17, 8 and 5 Gyr). For some given age and metallicity
combinations, a new population could match the old one in colour. The
lower left panel shows the additional constraint in order to also match 
the luminosity function of the new and old globular clusters. 
The dashed part of the isochrones represent age
and metallicity combinations for which the new population would differ
by more than 0.4 mag from the old one and therefore be distinguishable
in the luminosity function. Finally, the lower right panel combines all
constraints and shows as a shaded area the range of allowed age and metallicity
combinations for new globular clusters in order to look similar both in colour 
and magnitude to an old one. See text for further details.
}
\end {figure*}

Note that we have assumed above an approximately equal number of young and old 
globular clusters. The parameter range would be relaxed if the ratio of 
new to old (or vis--versa) globular clusters is larger
than 3:1, since it would be more difficult to disentangle the
two populations in the colour or magnitude distributions. The Milky Way might
be representative, with a halo and bulge/disk population in the ratio
of 4:1 (Minniti 1995). These populations have
roughly the same age, 
differ in mean metallicity by [Fe/H]$\simeq$ 1 dex but cannot 
be disentangled in a colour distribution (the dispersion in V--I lies
around $\sigma = 0.08$ mag). However, if we do assume an unbalanced
ratio of new to old globular clusters after a merger event, then, given
the observational facts of Sect.~2.1, this would suggest that different 
progenitors than gas--rich galaxies were involved in the merger, 
e.g.~at least one galaxy must be gas--poor to explain the low number of
newly formed globular clusters.

\subsubsection{Do low--luminosity ellipticals have two populations of
globular clusters?}

To summarize, the progenitor galaxies of a recent
merger must have had a significant globular cluster system; further,
observations suggest that a gas--rich merger will have induced the
formation of a large number of new globular globular clusters. The
fast and preferential destruction of one of these populations is
very unlikely. Therefore, the fact that we do not detect a second population
in low--luminosity ellipticals forces us to reject the assumption that these
galaxies formed in a gas--rich merger, or as discussed above, to assume 
that mean age
and metallicity of the new globular cluster population tightly conspire
to make the new population appear similar to the old one both in the
colour and magnitude distribution. However, in the latter case, 
the current photometric data imply mean ages
for the new population less than half the age of the old population 
are excluded (i.e.~cannot be compensated by any metallicity). This therefore
rules out that low--luminosity ellipticals formed by a {\it recent}
($z<1$) gas--rich merger.


\subsection{Two examples: NGC 1380 and NGC 1427}

In order to better illustrate the age--metallicity conspiracy, we briefly 
discuss two examples. The first
is NGC 1380, a relatively luminous (M$_V\simeq -21.5$) early--type (S0) galaxy 
which has sub--populations of globular clusters and
illustrates the accuracy with which age and metallicity
can be disentangled from broad--band colours. The second is NGC 1427, a 
low--luminosity elliptical (M$_V\simeq -20.5$), in which 
no sub--populations could be detected, and
which illustrates the current status of research on globular clusters in
low--luminosity galaxies.

NGC 1380 is the second brightest early--type galaxy in the Fornax
cluster, and its globular cluster system was studied with deep photometry 
in three broad--band
filters (Kissler-Patig et al.~1997b). Although the system is not very rich
($\sim 550$ globular clusters), it showed a bimodal colour distribution. 
The globular cluster luminosity functions of the two
sub--populations were indistinguishable. This could correspond in Fig.~1
to a second population of 12 Gyr with solar metallicity:
indistinguishable in magnitude (solid line), but different enough in
colour to be detected. The conclusion of that study was that the age 
difference between the 
two sub--populations must be $\simeq 4$ Gyr or less, and the colour
difference is mainly due to a metallicity difference (estimated to be
around 1 dex).  This example
illustrates perhaps the highest accuracy with which age and metallicity 
differences can be determined. The high sensitivity to sub--populations was
mainly due to deep photometry (complete down to B\ $=25.8$), the
associated small photometric errors, and the sensitivity to metallicity of the 
three combined colours (B, V, and R). We expect, with data of similar
quality, to be 
able to detect similar sub--populations in low--luminosity ellipticals.

Probably the best photometrically studied {\it low--luminosity}
elliptical is NGC 1427, an E3
galaxy in Fornax, a little more than a magnitude fainter than NGC 1380. 
It was studied by Forbes et al.~(1996) with HST, and Kissler-Patig et 
al.~(1997a) from the ground. We combined the ground--based data down to
$V<23.0$, with the HST data for $V>23.0$. The sample now includes 176
globular clusters with good V and I photometry (mean error in
V--I\ $= 0.1$ mag). The sample represents over 50\% of the total existing 
$320\pm60$ globular clusters within 3 r$_{\rm eff}$ of the galaxy.
No systematic differences were found between the photometry of the two data sets, 
thus no correction was necessary before combining the
samples. The globular cluster colour distribution is shown in Fig.~2.
No multiple populations are seen by eye, nor detected by a KMM (Ashman, Bird
\& Zepf 1994) statistical test.  We over--plot
a Gaussian with a width corresponding to our median error in V--I,  
$\sigma_{\rm \small ERR}=0.11$ mag, to illustrate
the broadening expected from the errors in the photometry alone. A free
Gaussian fit to the histogram gives a width of $\sigma_{\rm \small TOTAL}=0.15$
mag, thus the intrinsic width of the distribution must be less than 
$\sigma$(V--I)$=0.10$ mag.  
\begin{figure}
\psfig{figure=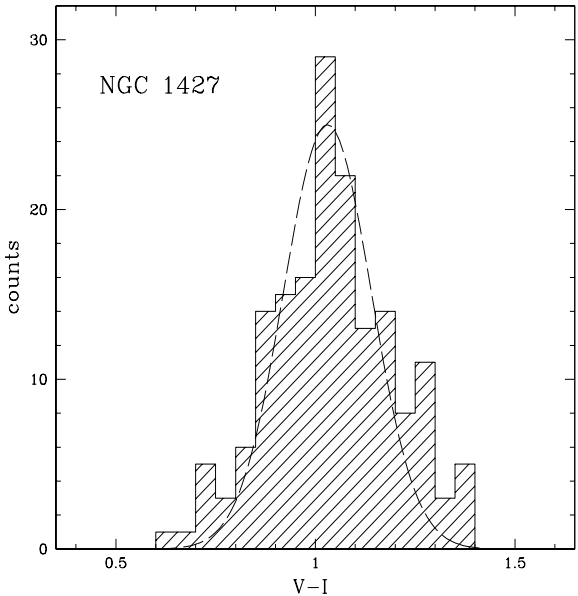,height=8cm,width=8cm
,bbllx=8mm,bblly=57mm,bburx=205mm,bbury=245mm}
\caption{Globular cluster colour distribution in NGC 1427 (M$_V = -20.5$).
The Gaussian
represents the broadening from the photometry errors alone. There is
no evidence for multi--modality. 
}
\end {figure}
According to the new V--I versus metallicity relation from spectroscopy of
NGC 1399 globular clusters (Kissler-Patig et al.~1998), the mean colour
reflects a metallicity of [Fe/H]$=-1.1\pm0.3$ dex, assuming an old mean age
(i.e.~similar to the age of the Milky Way globular clusters).

The luminosity function does not show any peculiarity either. It peaks in
V and I at the same value, within the error, as do the globular cluster
luminosity functions of other early--type galaxies in Fornax 
(Kohle et al.~1996). Using
an independent distance modulus of 31.4 (from Cepheids, Madore et al.~1997), 
the globular cluster luminosity function of NGC 1427 peaks at 
$M_V^{TO}=-7.6\pm0.3$ mag, as does the one for the Milky Way globular
clusters, supporting similar old ages as in the Milky Way. We conclude that 
NGC 1427 hosts a population of intermediate
metallicity, old globular clusters, and that no other population is visible.
Although V--I is not as sensitive as B--V and B--R (as was available for
NGC 1380) to metallicity, we can further
conclude that no sub--population of less than half the age of the old one
is present (see Sect.~3.2). Therefore NGC 1427 cannot have been
formed by a recent ($z<1$) dissipational merger. 

\subsection{High--luminosity ellipticals}

The above sections makes it clear that constraints can be put on the merger
history of low--luminosity ellipticals from their globular cluster
systems. Can similar constraints be applied to high--luminosity galaxies?
The answer seems to be no, because large ellipticals often show bimodal globular
cluster colour distributions (e.g.~compilation by Forbes, Brodie \&
Grillmair 1997), whereas the constraints derived above mainly come from
the detection of only one population. Moreover, the arguments
used for low--luminosity ellipticals turns up some contradictions
when applied to high--luminosity ellipticals.

High--luminosity galaxies are generally
thought to be the result of a merger involving little gas (see Sect.~2.2)
and are, therefore, expected to have formed only a few new 
globular clusters in interactions (see Sect.~2.1). But the observed colour 
distributions are often bimodal with an equal 
number of blue and red globular clusters within a factor two.
We are apparently facing two 
contradictions. The first being the fact that if the ages and metallicities
of globular clusters in small ellipticals conspire to give a unimodal 
distribution, it is
not the case in large ellipticals. The second is that the number of red
(assumed young) globular clusters in large ellipticals is roughly the same
as the number of the blue (old) globular clusters, in contradiction with the 
small amount of gas expected in the merger that formed the
high--luminosity ellipticals. Is there an explanation other than 
relaxing one of our conclusions of Sect.~2, or the assumption that the
red globular clusters in high--luminosity ellipticals formed in merger
events?

\subsubsection{No conspiracy in high--luminosity ellipticals?}

The first apparent contradiction can be explained. 
We do not necessarily expect a conspiracy between the age and metallicity
of the new globular clusters in high--luminosity ellipticals if
present in low--luminosity galaxies. Forbes, Brodie \& Grillmair (1997)
showed that the metallicity of the red globular clusters correlates with
the luminosity of the parent galaxy. A similar behaviour is known for
the stellar component of galaxies (the Mg--$\sigma$ relation,
e.g.~Burstein et al.~1988), and a similar relation for globular clusters
would not be unexpected if the red globular cluster
population is associated with the stellar light (e.g.~Kissler-Patig et
al.~1997b, Lee, Kim \& Geisler 1998). According to this relation, red globular
clusters in a low--luminosity galaxy (e.g.~M$_V=-20.0$) would 
have a metallicity lower by about [Fe/H]$=0.5$ dex than red globular clusters
in a high--luminosity galaxy (e.g.~M$_V=-22.5$). Assuming that the
metallicity of the blue populations in these galaxies are independent of
M$_V$ (no correlation with M$_V$
was found by Forbes, Brodie \& Grillmair 1997), a red population in a
high--luminosity galaxy would differ from the blue one while this might
not be the case in a low--luminosity galaxy. As an example, we assume
an old population similar to the one in Fig.~1. In a low--luminosity
galaxy (M$_V=-20.0$, similar to NGC 1427) a new population of 8
Gyr with a mean metallicity of [Fe/H]$=-0.5$ dex would be hidden in both
the colour and magnitude distribution of the globular cluster system.
The same new population would have a mean metallicity of
[Fe/H]$=0.0$ dex (solar) in a high--luminosity galaxy (M$_V=-22.5$,
similar to M87) and would show up in the colour distribution as a
second peak, offset by 0.2 mag in V--I from the old one. Note that this
new population would still be hidden in the globular cluster luminosity
function. This scenario is actually very close to the current
observations, e.g.~in M87, Whitmore et al.~(1995) and Elson \&
Santiago (1996) see two peaks in the colour distribution separated by
V--I$\simeq0.2$ mag and no significant ($\Delta V < 0.4$ mag) effect in 
the globular cluster luminosity function.

\subsubsection{Did the red globular clusters form in a merger event?}

The second contradiction is more serious. Given the observations
presented in Sect.~2.1, the large number of red globular clusters in
high--luminosity ellipticals is not compatible with a formation via
gas--poor mergers. Either high--luminosity ellipticals formed by
gas--rich mergers (but see Sect.~2.2), or the large number of red
globular clusters is not associated with merger events. We investigate
the latter possibility.

Recent spectroscopy of
blue and red globular clusters in NGC 1399 and M87 (Kissler-Patig et
al.~1998, Cohen, Blakeslee \& Rhyzov 1998)
has shown, that none of the major sub--populations has formed in a
{\it recent} ($z<1$) merger. Only a small fraction of very red objects in NGC
1399 seem
compatible with a more recent formation, e.g.~in a merger event
involving a moderate amount of gas. This
picture is supported to some extend by the photometry of globular clusters
in NGC 1380 (Kissler-Patig et al.~1997b). The origin of blue and red clusters
in this galaxy is unclear, both populations are old and could as well be
associated with the bulge/disk and halo formation, as with a merger
origin. Clearly the picture of  
blue globular clusters in a bimodal distribution being old and the red ones
being young and formed in a merger is too simple and misleading. 

We note that this prohibits a straightforward 
comparison of the properties of sub--populations with the expectations from
the merger scenario of Ashman \& Zepf (1992), as long as it remains unclear 
what fraction of the red globular clusters formed in a merger event. 
If other processes can form red globular
clusters, the number of red versus blue clusters does not support nor
rule out any merger model. 

Alternatives for forming the large number of globular clusters around
high--luminosity ellipticals and explaining the presence of sub--populations
were discussed by various authors. 
Formation of the red clusters in dissipational mergers {\it at a very
early time} in the history of the galaxy is not, a priori, 
excluded (although see arguments by Forbes, Brodie \&
Grillmair 1997).  Blakeslee (1996) and Blakeslee et al.~(1997) 
associate over--abundant globular cluster systems with 
the preferential position of the galaxy in a 
clusters, and with an early--formation of a 
number of globular clusters proportional
to the available mass. However they do not comment on the origin of the
different globular cluster sub--populations.
Forbes, Brodie \& Grillmair (1997) argued that the
two--populations in large ellipticals are the product of a two--phase
collapse. Kissler-Patig (1997b) suggested a global pre--galactic formation
in fragments and a formation in the collapse of the disk/bulge forming
the galaxies. The last two
scenarios differ in the sense that the former assumes rather
monolithic collapses with the metal--rich globular clusters in
ellipticals being analogous to the metal--poor ones in spirals, invoking
a third phase in late--type galaxies. The Kissler-Patig 
scenario would associate metal--poor globular
clusters in ellipticals with metal--poor globular clusters in spirals and
the intra--cluster medium. The metal--rich globular clusters in
ellipticals would be analogous to the disk/bulge populations in spirals. 
The common point,
independent of the exact formation mechanism, is that the sub--populations 
differ slightly in age and could differ considerably in metallicity
(depending on the details of the enrichment process), 
since the second generation will form from more enriched gas. 
But in both scenarios the sub--populations are intrinsic to the galaxy
and not mainly the product of a late merger. 

These scenarios have gained some support recently. Harris et al.~(1997)
ruled out large age differences between halo globular clusters in the Milky
Way. They interpreted their results in a picture where halo globular clusters
formed in fragments that started star formation within the same Gyr and 
later merged to build the Galaxy. Further, Elson (1997) has detected a blue
and a red stellar population in the early--type galaxy NGC 3115, with
typical halo and bulge metallicities, hinting at the fact that all galaxies
could have stellar halos but ellipticals are ``bulge'' dominated. The
recent findings in the early--type galaxies NGC 1380, NGC 1399 and NGC 4472 
that associate blue
and red globular clusters with halo and bulge/disk stellar populations support
this picture. As pointed out by Harris (1998) the early star formation in
fragments will presumably cause its own demise, and lead to a ``dormant''
phase as suggested by Forbes, Brodie \& Grillmair (1997). Star and globular
cluster formation would then restart during the collapse of the fragments
to a bulge/disk. We would then expect the presence of old, blue and red, globular
clusters in all galaxies, independently of their types. Note that such a
scenario would not rule out a contribution through mergers to the red
globular cluster population (e.g.~the very red globular clusters in NGC
1399, Kissler-Patig et al.~1998). 

Clearly, a definitive understanding of globular cluster 
systems requires accurate age, metallicity and kinematic determinations of 
the different sub--populations from an observational point of view, and
better predictions of the relative importance of the different globular cluster
formation mechanisms from a theoretical point of view.  
We conclude that the large number of red globular clusters in
high--luminosity galaxies imply that most did not form in a meger,
unless these galaxies did experience a number of dissipational merger events
in their past.

\subsection{The smoking guns}

The properties  of globular clusters in low--luminosity galaxies have left us
with two alternatives: these galaxies did not form in a gas--rich merger
event, or the age and metallicity of the new globular clusters conspire to
look similar in their colour and magnitude distribution to the old ones.
Can we discriminate between these two possibilities by investigating 
current day gas--rich mergers? In particular, 
what will these mergers and their globular cluster systems
look like after several Gyr? Will they resemble low--luminosity
galaxies such as NGC 1427?

The ongoing mergers listed in Sect.~2.1
typically involve gaseous sub--L$^{\ast}$ galaxies and therefore would be 
expected to form intermediate--luminosity ellipticals. For
example, Whitmore et al.~(1997) estimate the final (after several Gyr)
magnitude of NGC 1700 and NGC 3610 to be M$_V$ = --21.26 and --20.45
respectively. 
If it could be shown that the age and metallicity of globular clusters will 
conspire in merger remnants to make them look
unimodal in colour, and also if they match the other characteristics of
globular cluster systems in early--type galaxies (see
Sect.~3.2), then these systems might indeed be the progenitors of small, 
disky ellipticals. 
Whitmore et al.~(1997) simulated 
the evolution of the newly formed globular clusters with population synthesis
models. However, they did not allow for the metallicity of the new
globular clusters to vary (fixing it at a solar metallicity) which
excludes any age--metallicity conspiracy. 
They concluded that in this case, the newly formed globular clusters will 
look redder than the globular clusters of the progenitors 
after several Gyr, in agreement with our result in Sect.~3.2. Indeed,
for a large
metallicity difference, age will not be able to ``compensate'' the metallicity
effect on colour. The assumed
metallicity for the new globular clusters is critical for the 
colour distribution of the end--product, and before we can claim a
contradiction between ongoing mergers and low--luminosity galaxies, we
would need spectroscopic metallicity determinations of the new globular
clusters. 

A second aspect by which low--luminosity ellipticals and
merger remnants could differ, are the spatial distributions of their
globular cluster systems. The newly formed stars and
clusters in NGC 4038/4039 and NGC 1275 are preferentially located in the center,
and it remains unclear if the new systems will show the spatial
properties of today's low--luminosity globular cluster systems, namely
a surface density profile following that of the stellar light. This will 
depend on the amount and distribution of the newly formed stars.

In summary, it is unclear whether or not the ongoing mergers seen today and
their globular cluster systems will evolve into low--luminosity galaxies
as we see them, given the few observational constraints and in
particular the lack of information about the nature of the new clusters.


\section{Concluding remarks}

\subsection{Summary and present conclusions}

Recent observations of
merging galaxies indicate that globular clusters form in all gaseous
merger events, and the more gas is involved, the more new globular clusters
form. The observational data suggests that $\sim$10\% to $\sim$100\% new 
globular clusters are created compared to old ones when going from
early--type/early--type, to early--type/late--type, to
late--type/late--type
galaxy collisions. These observations naturally imply that 
low--luminosity ellipticals, thought to have formed in gaseous mergers,
should have two distinct populations of globular 
clusters -- old globular clusters from the progenitor galaxies and new 
ones created from the gas of the merger.
Observations of globular cluster systems in low--luminosity 
ellipticals to date do {\it not} show evidence for
two globular cluster populations. 
We have examined several solutions to this problem (Sect.~3.2).

\begin{itemize}
\item{The absence of sub--populations due to the absence of an old
blue population implies the early formation by merging essentially gaseous
progenitors that did not have time to build up an old globular cluster
system. Alternatively old globular clusters could have been preferentially
removed or destroyed. In both cases today's
globular clusters would be the ``new'' ones, but are observed to be as
old as the Milky Way ones.}
\item{The preferential destruction of the newly formed globular clusters 
is not ruled out, but we still see young clusters in mergers that are several
Gyr old. The absence of young clusters in low--luminosity ellipticals
would then set a lower age limit for the merger event.}
\item{If age and metallicity of young and old globular clusters tightly 
conspire to make them appear as one population, then constraints from 
the colour and magnitude
distributions of globular clusters allow us to constrain the maximum age and metallicity
differences between the two sub--populations to be about a factor of 
two in age, and 1 dex in [Fe/H].}
\item{If a large young population is absent because it simply did not form,
then the current observational evidence would rule out the formation of
low--luminosity ellipticals by gaseous mergers.}
\end{itemize}

All solutions to this problem, regardless of their likelihood,
imply that {\it low--luminosity ellipticals, if they formed by dissipational
mergers, formed at early times with a rough lower limit of $z>1$.}

While the absence of sub--populations in low--luminosity ellipticals can
put constraints on merger histories, the presence of sub--populations
in high--luminosity ellipticals is more complicated to interpret.
In particular, the multiple sub--populations are not likely to be 
the product of the last dissipationless merger event, and many plausible
alternatives to mergers exist to explain the formation of 
the globular cluster systems in these galaxies.

\subsection{Sharpening the constraints}

We have presented constraints on merger models from the properties
of their globular cluster systems in nearby ellipticals. 
Clearly, the study of globular clusters can
constrain these models, but evidently our knowledge is still very incomplete.
In the following we list briefly list the points that would help to
sharpen these constraints, and that were the limiting factors in our
analysis.
\begin{itemize}
\item{The nature of the young star clusters in ongoing mergers
needs to be better understood.
Will these objects really evolve into objects similar to old globular
clusters seen today and what is their metallicity?}
\item{The globular cluster systems of low--luminosity ellipticals need to 
be better studied with deep, accurate three--colour photometry,
or, even better, with multi--object spectroscopy to detect or finally rule out
the presence of sub--populations of globular clusters.}
\item{The origin of sub--populations needs to be better understood. Mergers
are one, {\it but not the only}, alternative for building up a second
population of globular clusters. Accurate, {\it unique}, predictions
for the different scenarios (mergers, in situ formation, stripping, accretion 
of dwarf galaxies, etc...) need to be explored via modeling, before the
properties of globular cluster systems can be fully interpreted.} 
\end{itemize}

\noindent
{\bf Acknowledgments}\\
We thank Jean Brodie for useful comments on the manuscript, as well as
the referee for a careful reading and helpful comments that substanstially
improved the manuscript.
MKP was partly supported by the {\sc DFG} through the Graduiertenkolleg
`The Magellanic System and other dwarf galaxies'. Part of this
research was funded by the HST grant GO.05990.01-94A. 
DM is supported by Lawrence Livermore National Laboratory, under DOE
contract W-7405-Eng-48.\\

\vskip 1cm

\noindent{\bf References}

\noindent
Ashman K.M., \& Zepf S.E. 1992, ApJ 384, 50 \\
Ashman K.M., Bird C.M., Zepf S.E. 1994, AJ 108, 2348\\
Ashman K.M., \& Zepf S.E. 1997, Globular Cluster Systems (Cambridge:
Cambridge University Press)  \\
Bender R., Surma P., D\"odereiner S., M\"ollenhoff C., Madejsky R. 1989,
A\&A 217, 35\\
Bender R., 1997, in ``The Nature of Elliptical Galaxies'', ASP Conf.Series
Vol.116, p.11, eds.~M.Arnaboldi, G.S.Da Costa, P.Saha \\
Blakeslee J.P. 1996, PhD thesis, MIT \\
Blakeslee J.P, Tonry J.L., Metzger, M.R. 1997, AJ 114, 482\\
Brodie J.P., Schroder L.L., Huchra J.P., Phillips A.C., Kissler-Patig M.,
   Forbes D.A. 1998, AJ submitted \\
Bruzual G., \& Charlot S. 1993, ApJ 405, 538 \\
Bruzual G., \& Charlot S. 1997, in preparation \\
Burstein D., Davies R.L., Dressler A., Faber S.M., Lynden-Bell D.,
Terlevich R., Wegner G. 1988, in ``Towards Understanding Galaxies at
Large Redshift'', eds. R.G.Kron \& A.Renzini (Dordrecht:Kluwer), 17\\
Chiosi C., Bressan A., Portinari L., Tantalo R. 1997, A\&A submitted\\
Cohen J.G., Blakeslee J.P. \& Rhyzov A. 1998, ApJ 496, 808\\
Dubath P., \& Grillmair C.J. 1997, A\&A 321, 379\\
Elson R.A.W., 1997, MNRAS 286, 771 \\
Elson R.A.W., Santiago B.X., 1996, MNRAS 278, 617\\
Elmegreen B.G., Efremov Y.N. 1997, ApJ 480, 235\\
Faber S.M., et al. 1997, AJ 114, 1771\\
Forbes D.A., Brodie J.P., \& Huchra J. 1996, AJ 112, 2448 \\
Forbes D.A., Franx M., Illingworth G.D., Carollo C.M. 1996, ApJ 467, 126\\
Forbes D.A., Brodie J.P., \& Grillmair, C.J. 1997, AJ 113, 1652 \\
Forbes D.A., Brodie J.P., \& Huchra J. 1997, AJ 113, 887 \\
Fritze v.~Alvensleben U., Gerhard O.E., 1994, A\&A 285, 775 \\
Fritze v.~Alvensleben U., Burkert A., 1995, A\&A 300, 58 \\
Geisler D., Lee M.G., Kim E., 1996, AJ 111, 1529 \\
Gnedin O.Y., Ostriker J.P., 1997, ApJ 474, 223\\
Harris W.E., 1991, ARA\&A 29, 543 \\
Harris W.E., 1994, in ``The Globular Cluster -- Galaxy Connection'' ASP
Conf.Series Vol.48, eds. G.H.Smith, J.P.Brodie \\
Harris W.E., 1996, in ``The Extragalactic Distance Scale'', STScI
Workshop \\
Harris W.E. 1998, in ``Galactic Halos'' ASP Conf.Series, ed.D.Zaritzky\\
Harris W.E., et al.~1997, AJ 114, 1030\\
Harris W.E., Harris G.L.H., McLaughlin D.E., 1998, AJ May issue\\
Hilker M., Kissler-Patig M., 1996 A\&A 314, 357 \\
Holtzman J.A., et al., 1992, AJ 103, 691 \\
Holtzman J.A., et al., 1996, AJ 112, 416 \\
Jong R.S.~de, Davies R.L., MNRAS 285, L1 \\
Kauffmann G., 1996, MNRAS 281, 487 \\
Kissler-Patig M., 1997a, A\&A 319, 83 \\
Kissler-Patig M., 1997b, PhD Thesis, Sternwarte Bonn\\
Kissler-Patig M., Kohle S., Hilker M., Richtler T., Infante, L., Quintana
H., 1997a, A\&A 319, 470\\
Kissler-Patig M., Richtler T., Storm M., Della Valle M., 1997b A\&A 327,
503\\
Kissler-Patig M., Brodie J.P., Forbes D.A., Grillmair C.J., Huchra J.A.
1998, AJ 115, 105\\
Kohle, S., Kissler-Patig M., Hilker M., Richtler T., Infante L., Quintana
H. 1996, A\&A 309, L39\\
Lee M.G., Geisler D., 1993, AJ 106, 493 \\
Lee M.G., Kim E., Geisler D., 1998, AJ 115, 947\\
Lutz D.,  1991, A\&A 245, 31 \\
Madore B.F., et al. 1997, Nature submitted\\
Meurer G.R., 1995, Nature 375, 742 \\
Minniti D., 1995, AJ 109, 1663 \\
Minniti D., Alonso V., Goudfrooij P., Jablonka P., Meylan G. 1996, ApJ 467, 221  \\
Muzzio J.C., 1987, PASP 99, 245 \\
Neilsen E.H. Jr., Tsvetanov Z.I., Ford H.C., ApJ 483, 745\\ 
N{\o}gaard--Nielsen H.U., Goudfrooij P., Joergensen H.E., Hansen L., 1993,
A\&A 279, 61 \\
Padoan P., Nordlund A.P., Jones B.J.T. 1997, MNRAS 288, 145\\
Renzini A., 1997, ApJ 488, 35\\
Richtler T., 1995, in ``Reviews of Modern Astronomy'', Vol.8,
eds.~G.~Klare, Springer, p.163 \\
Schweizer F., Seitzer P., 1992, AJ 104, 1039 \\
Schweizer F., Seitzer P., 1993, ApJ 417, L29 \\
Schweizer F., Miller B.W., Whitmore B.C., Fall S.M., 1996 AJ 112, 1839 \\
Van den Bergh S., 1990, in Dynamics and Interactions of Galaxies,
ed.~R.Wielen (Berlin:Springer), 492 \\
Whitmore B.C., Schweizer F., Leitherer C., Borne K., Robert C., 1993,
AJ 106, 1354 \\
Whitmore B.C., Schweizer F., 1995, AJ 109, 960 \\
Whitmore B.C., Sparks W.B., Lucas R.A., Machetto F.D., Biretta J.A.,
1995, ApJ 454, L73\\
Whitmore B.C., Miller B.W., Schweizer F., Fall M. 1997, 114, 797 \\
Worthey G., 1994, ApJS 95, 107 \\
Zepf S.E., Ashman K.M., 1993, MNRAS 264, 611 \\
Zepf S.E., et al., 1997, in preparation \\

%

\end{document}